# 证券成交量/价的行为是否像是一种几率波？

石磊磊 [a, b, c]

（2005 年 10 月 17 日）

[a]（北京师范大学管理学院系统科学学系，北京 100875）
[b]（中国科学技术大学近代物理系，合肥 230026）
[c]（中意人寿保险有限公司北京分公司代理人，北京 100738）
E-mail: Shileilei8@yahoo.com.cn 或 leilei.shi@hotmail.com

**摘要:** 考虑到在股票市场中成交金额对成交量和价格波动存在某种约束关系，本文作者通过成交金额来研究成交量与价格波动之间的相互关系。我们发现：随着交易时间的延长，累计交易量在交易价格区间逐渐地在成交价格平均值附近呈现峰化的分布特征。这一特征与体系在此间交易价格涨落的路径、时间序列或总成交量的大小无关。为了解释这种量价行为，我们运用物理学的方法，提出了成交能量关系假说，推导出一个不显含时间变量的证券成交量/价几率波方程，并且得到两组解析的成交量随价格变化的分布函数。通过实证检验，我们证明了在股票市场中存在着相干特性，初步验证了该模型的有效性。成交量价的行为像是一种几率波。这样，我们试图提出一个适用于描述金融交易市场的、微观和动态的成交价格波动几率的统一理论。

# Does Security Transaction Volume-Price Behavior Resemble A Probability Wave?

Leilei Shi [a,b,c]

(October 17, 2005)

[a] *Department of Systems Science, School of Management, Beijing Normal University, Beijing 100875, China*
[b] *Department of Modern Physics, University of Science and Technology of China, Hefei 230026, China*
[c] *Agents, Generali-China Life Insurance Co. Ltd. (Beijing Branch), Beijing 100738, China*
E-mail addresses: Shileilei8@yahoo.com.cn, or, leilei.shi@hotmail.com

**Abstract**

Motivated by how transaction amount constrain trading volume and price volatility in stock market, we, in this paper, study the relation between volume and price if amount of transaction is given. We find that accumulative trading volume gradually emerges a kurtosis near the price mean value over a trading price range when it takes a longer trading time, regardless of actual price fluctuation path, time series, or total transaction volume in the time interval. To explain the volume-price behavior, we, in terms of physics, propose a transaction energy hypothesis, derive a time-independent transaction volume-price probability wave equation, and get two sets of analytical volume distribution eigenfunctions over a trading price range. By empiric test, we show the existence of coherence in stock market and demonstrate the model validation at this early stage. The volume-price behaves like a probability wave. Thus, we attempt to offer a unified, micro, and dynamic wave theory on price volatility probability in financial market.

## 1. Introduction

Although there are many trading price models in financial market, none of them has the explicit price formation mechanism that is expressed by an analytical expression. Fama [1] and Ross [2] launched efficient market hypothesis and arbitrage pricing theory, respectively, based on rational trading assumption. Black and Scholes [3], together with Merton [4], derived a Black-Scholes-Merton model in terms of Samuelson's log-normal process or economic Brownian motion [5] that could be traced to Bachelier's dissertation regarding an option pricing problem [6]. In addition, Engle [7] formulated ARCH model, which was later developed to GARCH model by Bollerslev [8], to estimate price volatility error or nonlinear item. In recent years, some econophysicists begin using formulation in physics to develop asset pricing models in financial market. For example, McCauley and Gunaratne [9] showed how the Fokker-Plank formulation of fluctuations can be used with a local volatility to generate an exponential distribution for asset return.

In the past 20 years, there was a growing body of research studying price and volume. Gallant et al. [10] undertook a comprehensive investigation of price and volume co-movement using daily New York Stock Exchange data from 1928 to 1987. Gervais et al. [11] claimed the existence of high-volume return premium in stock market. Moreover, Zhang [12], an econophysicist, presented an argument for a square-root relationship between price changes and demand. Over the last few years, spin models are used in studying price and volume as the most popular models in econophysics [13]. Plerou et al. [14] applied a spin model and empirically addressed how stock prices respond to changes in demand. They found that large price fluctuations occur when demand is very small.

Ausloos and Ivanova [15] studied price and volume by introducing the notion of a generalized kinetic energy. A generalized momentum is also borrowed from classical mechanics. It is defined as the product of normalized transaction volume times the average rate of price change during a price moving average period. They emphasized at the close that these concepts might also serve in a dynamic equation framework. Wang and Pandey [16] followed the same terminology with somewhat different definitions. They defined trading momentum as the product of relative price velocity and a time-dependent "mass", a normalized trading volume in a time interval, i.e. the volume liquidity. However, traditional literature on price and volume mainly focuses on the correlation between return and total volume (over a trading price range) in a given time interval.

Some scholars attempted to explain the behavior of price and volume. Admati and Pfleiderer [17] developed a theory in which concentrated-trading patterns arise endogenously as a result of the strategic behavior of liquidity traders and informed traders. Wang [18] used ICAPM to establish a theoretical links between prices and volume. Econophysicists, for example, Gabaix et al. [19] proposed a theory to provide a unified way to understand the power-law tailed distributions of return and volume, the non-normal distributions that have been caught much attention by econophysicists since Mandelbrot's finding [20]. But current theories credited the correlation between price and volume to a variety of factors, for examples, (optimal) trading motive and information quality etc. "What is surprising is how little we really know about trading volume" [21].

Soros [22] guessed: In natural sciences, the phenomenon most similar to that in financial economics probably exists in quantum physics, in which scientific observation generates Heisenberg's uncertainty principle…… Unfortunately, it is impossible for economics to become science……

Among econophysicists, however, Schaden [23] prudently discussed the possible generic aspects of quantum finance to model secondary financial markets and the challenges we probably had to face in this potential interdisciplinary field. Piotrowski and Sladkowski [24] published a series of papers on quantum finance and currently proposed the price model that uses complex amplitudes whose squared modules describe price movement probabilities, inspired by quantum mechanical evolution of physical particles. Kleinert [25] applied path integrals to the price fluctuation of assets, considering the prices as a function of time. Baaquie et al. [26] developed a derivative pricing model based on a Hamiltonian formulation, and wrote that it was too difficulty to solve Merton-Garman Hamiltonian analytically in most pricing problems [27]. Jimenez and Moya [28], on the other hand, showed that it is possible to obtain quantum mechanics principles using information and game theory etc. These researches are based on existed formulation and principle in quantum mechanics.

There is a celebrated dictum in Walls Street: Cash is king. Inspired by Soros's guess and motivated by how transaction amount constrain trading volume and price volatility in stock market, we study the relation between volume and price through amount of transaction in terms of physics.

Stock market appears a complex system because of a variety of interacted and coupled trading agents in this open and fully competitive market. Thus, it is probably a key for us to model it successfully how we observe the system and find a simplified methodology.

Price is volatile upward and downward to its mean value in intraday transactions on individual stock. We observe that accumulative trading volume gradually emerges a kurtosis near the price mean value over a trading price range when it takes a longer trading time, regardless of actual price fluctuation path, time series, or total transaction volume in the time interval [29,30]. Moreover, the volumes are not distributed normally. Whereas some of the distributions appear to be normal, others show to be wave, and the others exhibit to be exponent. These phenomena can not be explained by a current economic and finance mainstream theory—both a rational trading assumption and a price volatility random walk hypothesis.

Is the volume-price behavior driven by a kind of restoring force? Unlike previous literature, this paper investigates how total trading volume distributes over a trading price range in a given time interval and why it does in market dynamic perspective. Solving the volume-price probability wave equation that is derived from a transaction energy hypothesis rather than an existed formulation in finance or physics, we find two sets of analytical volume distribution eigenfunctions over a trading price range in a given time interval (volume liquidity distribution eigenfunctions). By empiric test, we demonstrate our model validation at this early stage.

The rest of the paper is organized as follows. In section 2, we use the absolute of zero-order Bessel eigenfunction model to fit and test the volume distributions over a trading price range and draw a major conclusion that our observation holds true; In order to explain our observation and volume-price behavior, we propose a transaction energy hypothesis and use a potential function to describe price volatility in Section 3; We, then, assume that there is the restoring force that drives trading toward an equilibrium price, and identify the linear potential (energy) that can represent for effective supply-demand quantity restoring force in stock market in Section 4; Section 5 constructs a Hamiltonian that is equal to the sum of transaction dynamic energy and potential energy, derives a time-independent security transaction volume-price probability wave equation, and gets two sets of analytical solutions; In Section 6 are analyses, discussions, and possible applications (practical uses); And final are summaries and conclusions.

## 2. Empirical results

In order to demonstrate that a stock total transaction shows a volume kurtosis near the price mean value over a trading price range in an intraday time interval, we test a group of individual stock daily trading data using econometric methodology. We collect the first 30 individual stock data samples in Shanghai 180 Index in June, 2003. Each sample is total transaction volume of a stock, which distributes over a trading price range, in an intraday trading time interval. There are total 630 ($30 \times 21$) samples. However, 11 samples are halt samples and one sample data is lost. Total tested samples are 618.

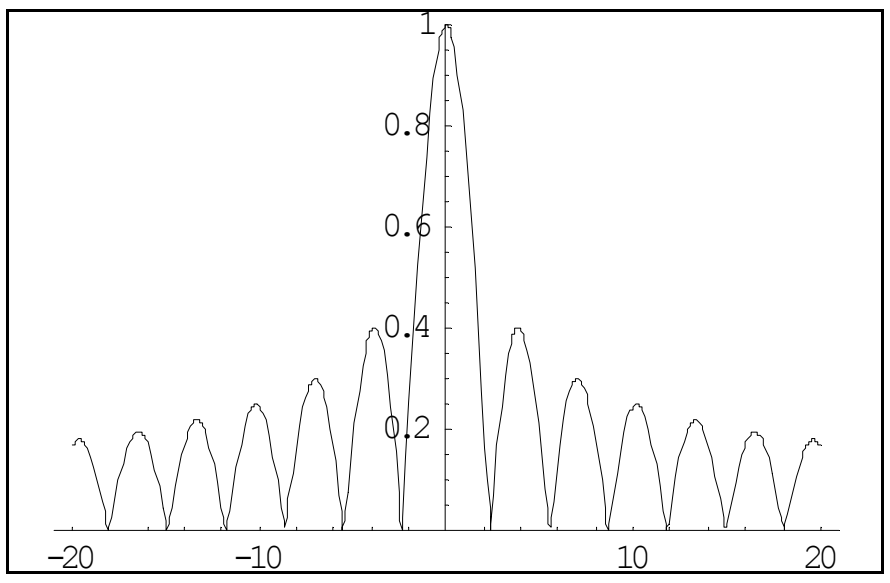

Fig. 1. The absolute zero-order Bessel Eigenfunctions

We use the absolute of zero-order Bessel eigenfunction as a test model (see fig. 1). It is

$$P_i(p_i) = C_m \left| J_0[\omega_m (p_i - p_0)] \right| + \varepsilon_i, \qquad (i = 1,2,3 \cdots N) \quad (1)$$

where $J_0[\omega_m (p_i - p_0)]$ is a zero-order Bessel eigenfunction, and its absolute normalized is a theoretical volume distribution probability at price $p_i$; $C_m$ is a normalized constant; $P_i(p_i)$ is an observed volume probability that is equal to the ratio of accumulative trading volume at price $p_i$ to total volume over a trading price range; $p_0$ is an equilibrium price if it exists; $\omega_m$ is an eigenvalue constant; And $\varepsilon_i$ is a set of random error [31].

Origin 6.0 Professional is friendly used in our test. Fig. 2 illustrates the fitting results of three typical samples. One appears to be normal, another shows to be wave, and the other exhibits to be exponent.

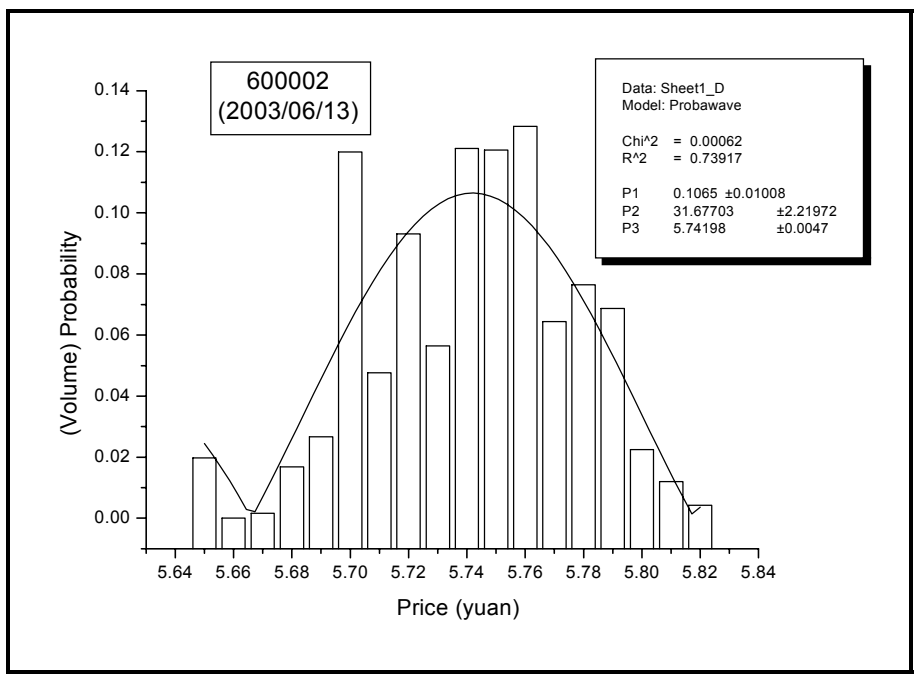

Fig. 2 (a). Fitting goodness in sample 1
(Its price mean value is RMB5.74 yuan)

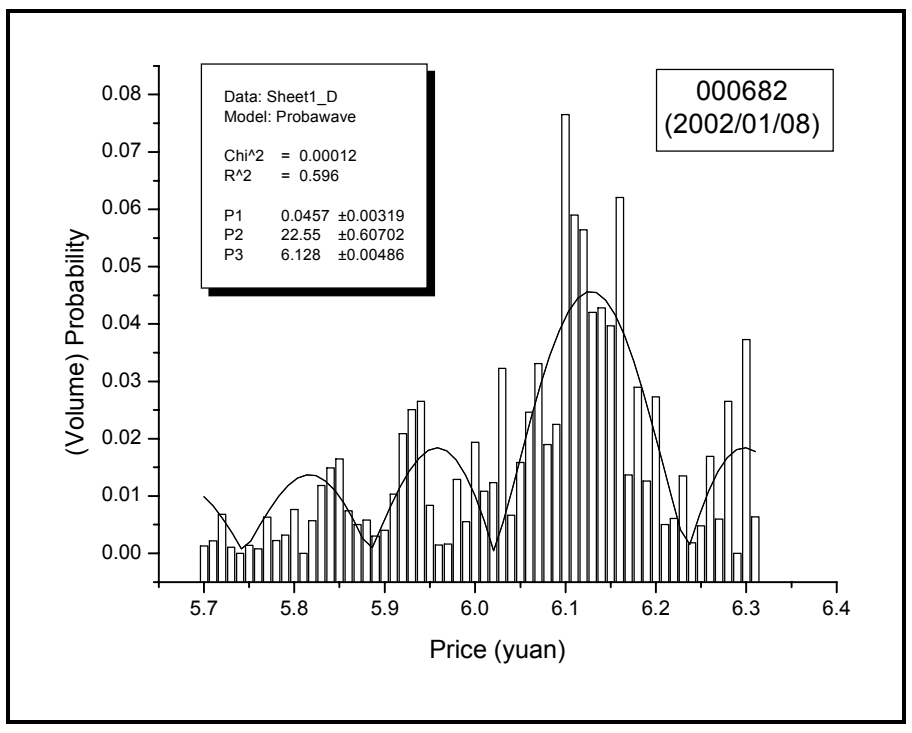

Fig. 2 (b). Fitting goodness in sample 2
(Its price mean value is RMB6.08 yuan)

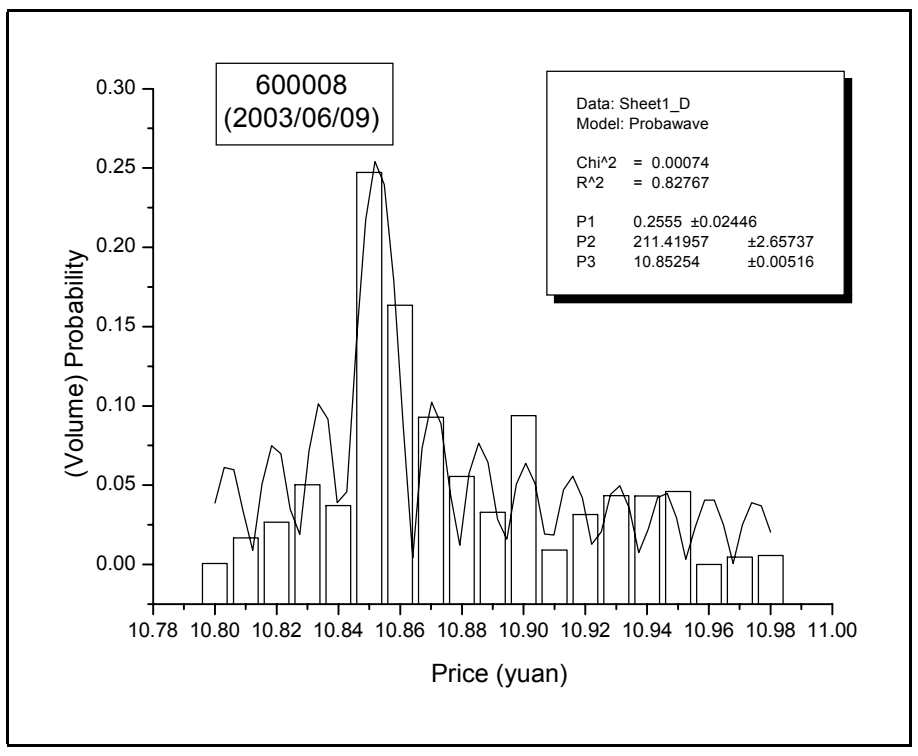

Fig. 2 (c). Fitting goodness in sample 3
(Its price mean value is RMB10.87 yuan)

The empiric test results that 94.34% of total samples shows significance at 95% level. The volume distribution has a main kurtosis over a trading price range in great majority situations. Thus, we conclude that our observation holds true.

Is there any a kind of restoring force that drives trading toward an equilibrium price to form those tested patterns? Does the volume-price behavior resemble a probability wave? What is the coherence in this economic system if it does? All of these questions are leading us to find a unified theory to reply and explain them.

## 3. Transaction energy hypothesis

In this section, we are going to establish a transaction energy hypothesis so that we can study the volume-price behavior in details.

The object of our study is accumulative trading volume (i.e. effective supply-demand quantity or transaction volume) distribution over a trading price range in intraday transaction on individual stocks. The reference or the origin of independent generalized coordinate is the price zero point. And, the benchmark is amount of transaction in a given time interval.

A transaction system is defined as a set of accumulative interacting trading in a given time interval, for example, a set of intraday transactions on a stock. In order to study collective interaction behavior in stock market, we define transaction volume $v$. It is accumulative trading volume or effective supply-demand quantity at a corresponding price $p$, if not specified. Its dimension is [share]. Total transaction volume $V$ is the sum of transaction volume over a trading price range. Its dimension is [share], too.

If transaction volume is fixed, but trading time interval is adjusted, then, there is a different effective supply-demand or trading activity in the market. The shorter it takes time, the more active the transaction is in the market. The balance between supply and demand is more difficult to be changed. The market is more stable. Therefore, we introduce transaction volume liquidity, which is defined as

$$v_t = \frac{v}{t}, \tag{2}$$

where $t$ is a trading time interval for the transaction volume, and its dimension is [time]; $v_t$ is transaction volume liquidity, and its dimension is [share][time]$^{-1}$.

Similarly, if transaction amount is fixed, but trading time interval is adjusted, then, there is different transaction amount liquidity in the market. The shorter it takes time, the better the transaction amount liquidity is in the market. The market is more stable. We introduce transaction amount in a given time interval or transaction amount liquidity to measure effective cash supply/demand liquidity in the market. It is

$$m_t = \frac{m}{t} = \frac{pv}{t} = pv_t, \tag{3}$$

where $p$ is price, and its dimension is [currency unit][share]$^{-1}$; $m$ is transaction amount or the product of transaction volume and price, and its dimension is [currency unit]; And $m_t$ is transaction amount liquidity, its dimension is [currency unit][time]$^{-1}$, and it is benchmark in this study.

In order to keep dimension consistence in derivation later, we introduce the notion of transaction energy. It is transaction amount liquidity rate or transaction amount acceleration at a price. It is defined as

$$E = \frac{m_t}{t} = \frac{pv_t}{t} = pv_{tt}, \tag{4}$$

where

$$v_{tt} = \frac{v_t}{t} = \frac{v}{t^2} \tag{5}$$

is transaction volume liquidity rate or transaction volume acceleration, and its dimension is [share][time]$^{-2}$; $E$ is transaction energy (cash liquidity energy) and its dimension is [currency unit][time]$^{-2}$.

The transaction volume probability is

$$P = \frac{v}{V}. \tag{6}$$

Let

$$E \equiv PE + (1-P)E \equiv T(p, v_t) + W(p, v_t)\,. \qquad (7)$$

where we define $T(p, v_t)$ and $W(p, v_t)$ are dynamic energy (distribution energy) and potential energy, respectively. Their dimension is [currency unit][time]$^{-2}$, too. Potential energy represents the interaction between the volume $v$ and the volume $V - v$ in a given time interval, namely, the interaction between the volume liquidity $v_t$ and the volume liquidity $V_t - v_t$.

When we focus on a stationary trading state, the potential is time independent. Thus, we write equation (7) as

$$-E + \frac{v_t^2}{V} p + W(p) = 0\,. \qquad (8)$$

Equation (8) is a transaction energy hypothesis in a differential expression. It is that transaction energy is equal to the sum of its dynamic energy and potential energy over a trading price range, which could be analogous to a conserved energy system in physics.

If $v / V = 1$, i.e. there is no price volatility in a given trading time interval, then, $W(p) = 0$; If $W(p) > 0$, then, $v / V < 1$, i.e. there is price volatility in the time interval. The potential energy can describe price volatility.

Equation (8) is a holonomic constraint (a generalized velocity independent constraint, e.g. $\Phi(p, v_t) = 0$) at price coordinate. According to classic dynamics [32,33], the number of degrees of freedom is equal to the number of generalized coordinates minus the number of independent equations of constraint. In our study, generalized coordinates are price and transaction volume in a given time interval (transaction volume liquidity). Thus, the number of independent generalized coordinates is one if we choose price as an independent generalized coordinate. When we do so, it is a one-dimensional problem describing the volume-price behavior as a whole.

## 4. Linear potential (energy)

In Section 3, we propose a transaction energy hypothesis and know that potential energy can be used in describing price volatility. Now, we assume that there is the restoring force that drives trading price toward an equilibrium price in a transaction system. Let us study how the force is produced in a financial economics perspective, and find what can represent for it in terms of a potential.

According to economic theory, buying or selling behavior is constrained. Demand quantity is constrained by budget or amount of the cash hold by buyers, while supply quantity is constrained by the volume of a product or a stock which is hold by sellers. Accumulative trading or transaction volume is constrained by both the price and amount of transactions. On the other hand, trading price and its volatility are constrained by amount of transactions and the volume.

How does the volume-price behave in a set of intraday transactions on an individual stock? Assume that a stock initial price is its equilibrium price, and total amount of cash used in buying stock (demand quantity) and total stock shares to be sold (supply quantity) are given.

Relative to present price, first, we assume that the price volatility is positive if demand quantity is greater than supply quantity, whereas it is negative if demand quantity is less than supply quantity.

Now, let us consider a restoring force in the system. Relative to initial price, if the balance between supply quantity and demand quantity is constantly maintained on intraday transactions, then, the trading price is its initial or equilibrium price. There is no price volatility at all (Here, the transaction potential energy is equal to zero in the time interval). Any a variable is exclusively determined by the other two among the volume, price, and amount of transaction.

Relative to an equilibrium price, supply quantity is frequently not equal to demand quantity because of separate individual buying and selling decisions in stock market. If demand quantity is greater than supply quantity in a short term interval, then, the trading price is volatile positively and deviates from its equilibrium price. The more the price deviates from it, the higher the price is. The same amount of remained cash could buy less volume of stock shares. The demand quantity is reduced. When the demand quantity is less than the supply quantity, the price will be volatile negatively and show reversal. It adjusts toward its equilibrium price. When reaching equilibrium price, the price volatility will continue in the same direction and deviate from equilibrium again because of price motion inertia. The more the price deviates from equilibrium, the lower the price is. The same amount of remained cash could buy more volume of stock shares. The demand quantity is increased. When the demand quantity is greater than the supply quantity once again, the price will be volatile in a positive direction and adjust toward its equilibrium point again until it returns to its equilibrium price once a more time. Hereafter, the price will repeatedly deviate from its equilibrium price and adjust toward it.

In a transaction system, trading price deviates from its equilibrium price because of quantity imbalance between supply and demand in a short term. The deviation changes the imbalance and causes the price reversal. When the price returns to its equilibrium price, it will continue its volatility in the same direction because of price motion inertia until it appears reversal again and then returns to its equilibrium price. It demonstrates that there exists an effective supply-demand quantity (i.e. transaction volume) restoring force that drives trading toward its equilibrium price in addition to generalized transaction force in a transaction system.

Because transaction potential energy is proportional to price, it is a linear potential. Moreover, the price moves upward and downward to its equilibrium price in a set of intraday transactions on individual stock. And, the volume-price behavior is driven by effective supply-demand quantity restoring force. Therefore, the linear potential that can represent for the restoring force is

$$W(p) = A(p - p_0) \approx A\left(p - \overline{p}\right), \tag{9}$$

where $p_0$ is an equilibrium price, $\overline{p}$ is its price mean value; $A$ is a transaction coefficient, and its dimension is [share][time]$^{-2}$.

Differentiating equation (8) by using equation (4) and (9), we define transaction restoring force quantitatively in this holonomic constraint system as

$$F_R = -\frac{\partial W}{\partial p} = -A = -\left(v_{tt} - \frac{v_t^2}{V}\right) = -\left(1 - \frac{v}{V}\right)v_{tt}, \tag{10}$$

where $F_R$ is restoring force. Thus, it is clear that coefficient $A$ is the magnitude of effective supply-demand quantity restoring force in a transaction system. The minus sign means that the force is always toward to its equilibrium price.

From the definition, equation (10), we can perceive that the restoring force is equal to zero if supply quantity and demand quantity is constantly balanced, i.e. $v / V = 1$ is satisfied. The linear potential can represent for an effective supply-demand restoring force in stock market.

**5. Probability wave equation and its solutions**

We have proposed a transaction energy hypothesis in Section 3 and identified the linear potential that can represent for effective supply-demand quantity restoring force in Section 4. Based on both, we will derive a differential equation describing the volume-price behavior and find its solution.

Suppose that a transaction volume-price probability wave function $\psi(p)$ is

$$\psi(p) = R \cdot e^{iS/B}, \tag{11}$$

where $R$ is the wave amplitude, $S$ its action or Hamiltonian principal function describing the volume-price behavior, and $B$ a constant to make its phase dimensionless [34,35].

From equation (11), we have

$$\frac{\partial S}{\partial p} = -\frac{iB}{\psi}\frac{\partial \psi}{\partial p}. \tag{12}$$

Assume that the volume-price behavior satisfies a Hamilton-Jacobi equation in this interacting energy system (see paragraph 5 in discussion 6.2.3. in Section 6) as

$$\frac{\partial S}{\partial t} + H(p, \frac{\partial S}{\partial p}) = 0, \tag{13}$$

where we construct the Hamiltonian that is equal to the sum of transaction dynamic energy and potential energy. It is written as

$$H(p,\frac{\partial S}{\partial p}) = T\left(p,\frac{\partial S}{\partial p}\right) + W(p)\,. \tag{14}$$

According to transaction energy hypothesis, equation (7) or (8), the transaction energy is always equal to the sum of transaction dynamic energy and potential energy over a trading price range no matter whether it is a constant. In comparison with transaction energy or potential energy, moreover, dynamic energy (distribution energy) is an error item when total transaction volume is large enough according to probability and statistics, and the error change could be ignored. On the other hand, when we study the error change, we could regard both transaction energy and potential energy as "infinite" constants. In addition, our major objective at initial derivation is to identify transaction momentum at price coordinate so that we can study its non-linear or second order item (dynamic or distribution energy) by building its second order differential equation in addition to studying price volatility linear item (potential energy)*. The momentum is supposed to be defined in this economic system no matter whether the system is conservative or dissipative. Thus, we could assume that both transaction energy and potential energy are constants at first in our derivation†. After we establish the (momentum second order differential) equation, we, then, study in details in this volume-price probability wave equation whether it is conservative or dissipative (please see its solutions later).

Now, we can find a special solution for equation (13) as follow:

$$S(p,t) = S_1(p) - Et\,, \tag{15}$$

or

$$\frac{\partial S}{\partial t} = -E\,, \tag{16}$$

where we suppose that the action in wave function (11) describe the volume-price behavior in a given time interval i.e. transaction amount liquidity behavior; $E$ is any time and price explicit independent energy (not necessary to be a constant over a trading price range), and its dimension is [currency unit][time]$^{-2}$.

Because the dimension of $S(p,t)$ or $S_1(p)$ is the same as that of $Et$ or $pv_t$, we can write the action as

$$S(p,t) = \alpha(pv_t) - Et + \beta\,, \tag{17}$$

where $\alpha$ is any a dimensionless constant and $\beta$ is any a constant with dimension [currency unit][time]$^{-1}$; In convenience, let $\alpha = 1$ and $\beta = 0$, then, we define

$$S \equiv pv_t - Et\,. \tag{18}$$

By equation (18), we critically define the generalized momentum or transaction momentum at price coordinate as

$$Q \equiv \frac{\partial S}{\partial p} = v_t\,. \tag{19}$$

According to action force definition in classical dynamics, we have transaction force $F_T$ as

$$F_T \equiv \frac{dQ}{dt} = \frac{dv_t}{dt} = v_{tt}\,. \tag{20}$$

Equation (20) shows that the force is transaction volume liquidity rate or transaction volume acceleration. Its demission is [share][time]$^{-2}$.

Substituting the momentum, equation (19), into equation (8), we write the transaction energy hypothesis or the Hamilton-Jacobi equation (13) as

$$-E+\frac{p}{V}\left(\frac{\partial S}{\partial p}\right)^2+W(p)=0\,. \qquad (21)$$

---

* In fact, Engle [7] estimates price volatility error or non-linear item by ARCH model. But econometrics does not have its analytical expression.

† Actually, we also assume that potential energy is a constant at first in the derivation of Schŏrdinger's equation although it is not in great majority cases in physics. We mainly focus on investigating a momentum second order item, kinetic energy, by building a second order differential equation as well [34,35].

We, then, construct a transaction energy functional $G(p,\psi)$ using equation (12) as

$$G(p,\psi)=(W-E)\psi^*\psi+\frac{B^2}{V}p\left(\frac{\partial\psi^*}{\partial p}\right)\left(\frac{\partial\psi}{\partial p}\right). \qquad (22)$$

According to security trading regulation, transaction priority is given by price first and time first regardless of information quality and participant's rationality. If current trading price is $p_c$, then, the next trading priority is offered to minimize the price volatility with respect to $p_c$. Therefore, we have its mathematic expression as

$$\delta\int G(\psi,p)dp=0\,. \qquad (23)$$

It is that actual trading price path is chosen by its energy functional to minimize its wave function $\psi$ with respect to price variations.

We, therefore, derive and write a time-independent transaction wave equation as

$$\frac{B^2}{V}\left(p\frac{d^2\psi}{dp^2}+\frac{d\psi}{dp}\right)+\left[E-A(p-p_0)\right]\psi=0\,, \qquad (24)$$

and its natural boundary conditions are

$\psi(0)=0$, $\psi(p_0)<\infty$, and $\psi(+\infty)\to 0$.

(Please see Appendix A for the derivation of a time-dependent equation)

To solve equation (24), we first choose $p_0$ as an origin in $o'-p'$ coordinate and a natural unit $\frac{V}{B^2}=1$. Obviously, the equation is supposed to keep its validation regardless of an origin selection. After solving the equation in a new coordinate, we let $p'=p-p_0$ and, then, get final solution for this problem.

We get two sets of analytical solutions from equation (24). If $E=p'v_{tt}$, then, we have a set of analytical solutions as

$$\psi_m(p)=C_mJ_0[\omega_m(p-p_0)]\,, \qquad (m=0,1,2\cdots) \qquad (25)$$

where $J_0[\omega_m(p-p_0)]$ are zero-order Bessel eigenfunctions, $C_m$ are normalized constants; and $\omega_m$ are eigenvalues ($\omega_m>0$) and satisfy

$$\omega_m^2=v_{tt}-A=F_T+F_R=\frac{v}{V}v_{tt}=const. \qquad (m=0,1,2\cdots) \qquad (26)$$

The absolute of function (25) is

$$\left|\psi_m(p)\right|=C_m\left|J_0[\omega_m(p-p_0)]\right|\,, \qquad (m=0,1,2\cdots) \qquad (27)$$

where $\left|\psi_m(p)\right|$ is transaction volume density or probability at price $p$ in a given time interval, i.e. the volume liquidity density or probability (see fig. 1). It has already been tested in Section 2 (see discussion 6.2.2. in Section 6).

If transaction energy is a constant over a trading price range (a conserved system in a physics perspective), then, we have the other set of analytical solutions from equation (24) as follows (see Appendix B):

$$\psi_m(p)=C_me^{-\sqrt{A_m}|p-p_0|}\cdot F\left(-m,1,2\sqrt{A_m}|p-p_0|\right), \qquad (m=0,1,2\cdots) \qquad (28)$$

with

$$\sqrt{A_m}=\frac{E_m}{1+2m}=const.>0\,, \qquad (m=0,1,2\cdots) \qquad (29)$$

where $F(-m,1,\sqrt{A_m}|p-p_0|)$ is a confluent hypergeometric function or the first Kummer's function; $A_m$ is the magnitude of restoring force or eigenvalue constant; $C_m$ are normalized constants; $m$ is the order of the eigenfunctions. The absolute of function (28) is

$$|\psi_m(p)| = C_m e^{-\sqrt{A_m}|p-p_0|} \cdot \left|F\left(-m,1,2\sqrt{A_m}|p-p_0|\right)\right|, \quad (m = 0,1,2\cdots) \quad (30)$$

where $|\psi_m(p)|$ is transaction volume probability at price $p$ in a given time interval.

If $m = 0$, then, $F\left(0,1,2\sqrt{A_0}|p-p_0|\right) \equiv 1$. The volume distribution $|\psi_0(p)|$ is exponent over a trading price range, which is sufficiently covered by function (27) (Compare Fig. 1 with Fig. 3 (a)); The eigenfunctions (30) are plotted in fig. 3, in which $m = 0,1,2 and 10$.

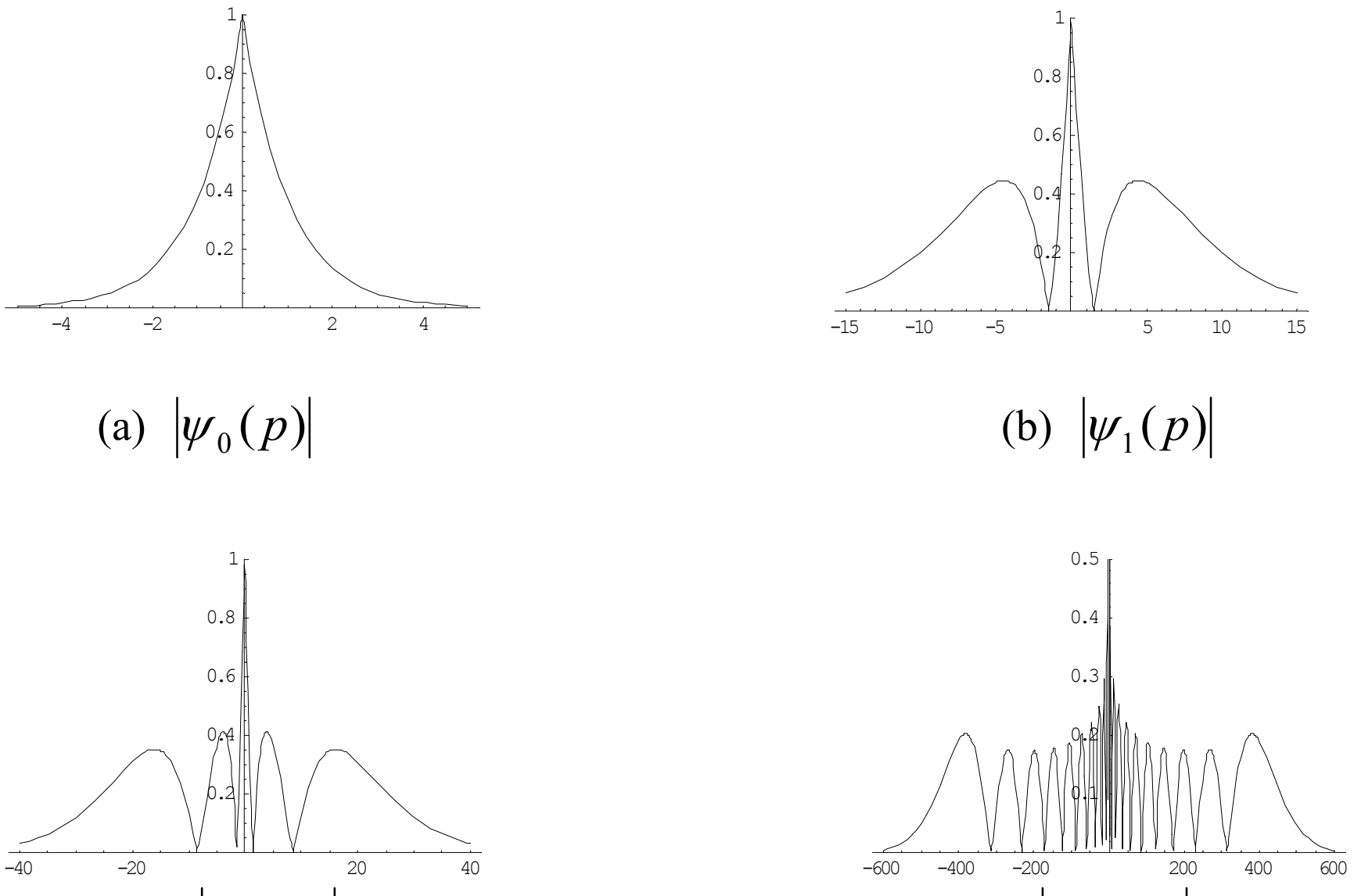

(a) $|\psi_0(p)|$ (b) $|\psi_1(p)|$

(c) $|\psi_2(p)|$ (d) $|\psi_{10}(p)|$

Fig. 3. The eigenfuntions if the magnitude of restoring force is a constant over a trading price range

## 6. **Analyses, Discussions, and Possible Applications**

In this section, we will have some analyses, discussions, and possible applications regarding our model.

### *6.1. Analyses on abnormal volume distribution patterns*

In our test, 35 samples among 618 lack significance at 95% level. They are characterized by at least two of trading volume kurtosis over a trading price range (see fig. 4). Let us explain it. Relative to an equilibrium price, the quantity imbalance between supply and demand has been increased too much so as to cause their equilibrium price to have a step or jump change in those trading samples. Price is volatile around from one equilibrium price to another. In such a situation, the volume distribution over a trading price range can be represented by the superposition of two eigenfunctions as follows:

$$|\psi_m(p)| = C_m\left(|J_0[\omega_m(p-p_{01})]| + |J_0[\omega_m(p-p_{02})]|\right). \quad (m = 0,1,2\cdots) \quad (31)$$

Fig. 4 shows the test results of two typical samples fitted by eigenfunctions (31). Among 35 samples, 34 samples have significance at 95% level.

This test characters an equilibrium price step or jump change from time to time. A transaction state still remains uncertainty if an eigenvalue is known (see Fig. 4 (a)).

The rest sample lacking significance is approximately a uniform distribution (see fig. 5). Fig. 5 and fig. 6 are the sample test results fitted by function (31) and the first-order function (30), respectively. If it is fitted by the first order eigenfunctions (30), then, the test shows significance at

95% level (see fig. 6).

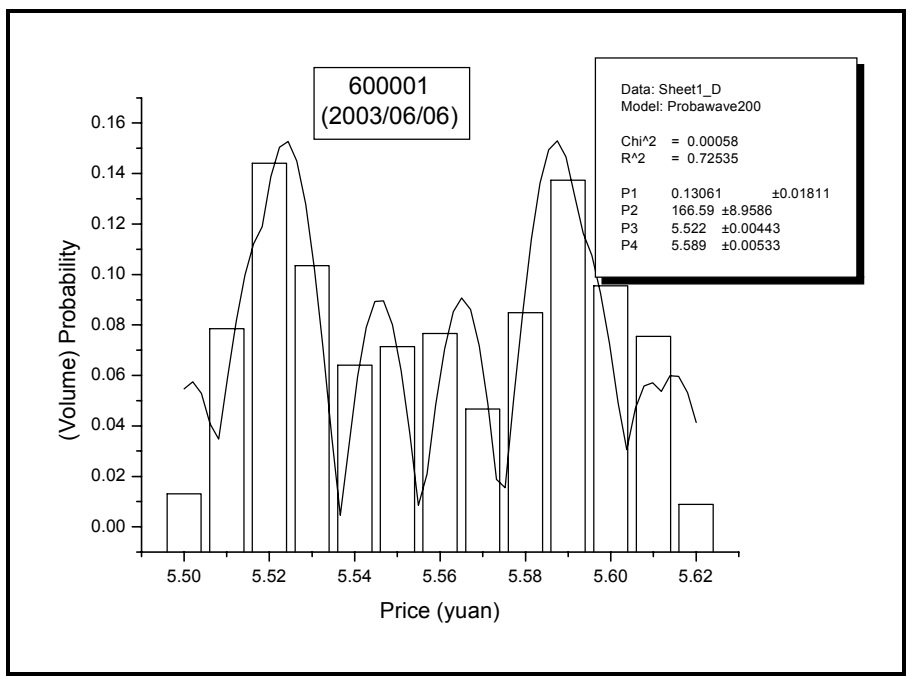

Fig. 4 (a). Fitting goodness by the superposition of two functions with an eigenvalue (31 among 34 samples)

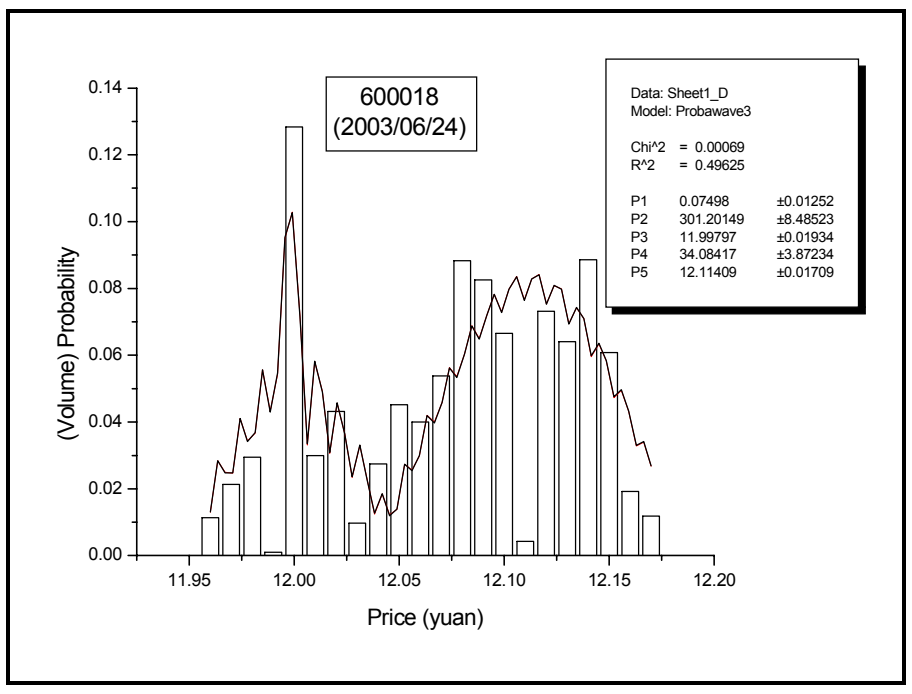

Fig. 4 (b). Fitting goodness by the superposition of two functions with two eigenvalues (3 among 34 samples)

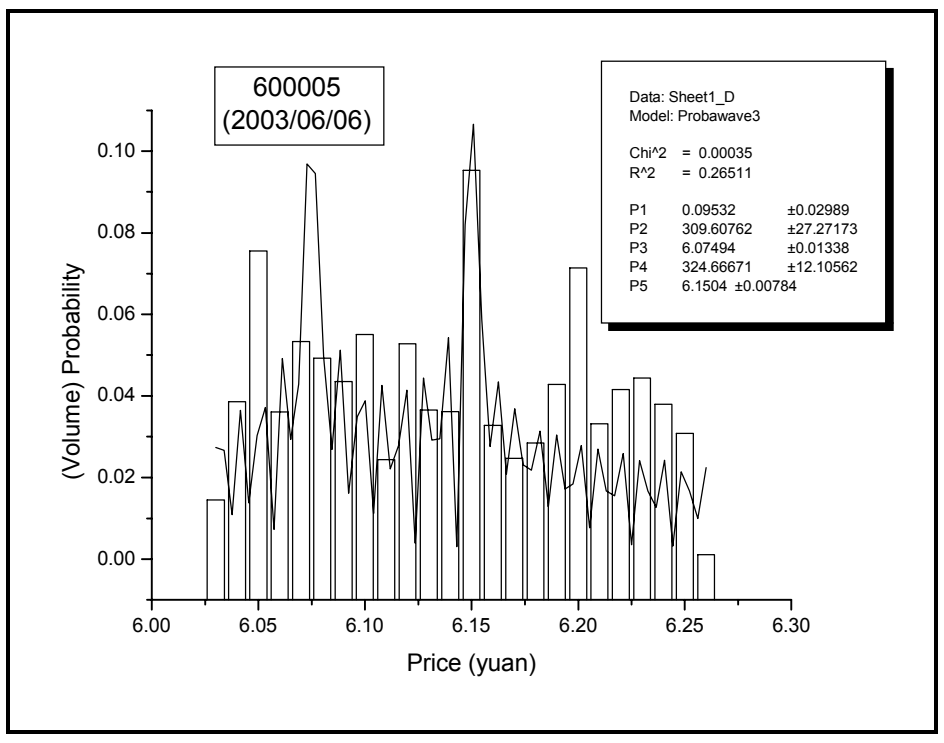

Fig. 5. The sample tested by superposition function (31) at 95% level ($R^2 = 0.27 < R^2_{crit} = 0.29$)

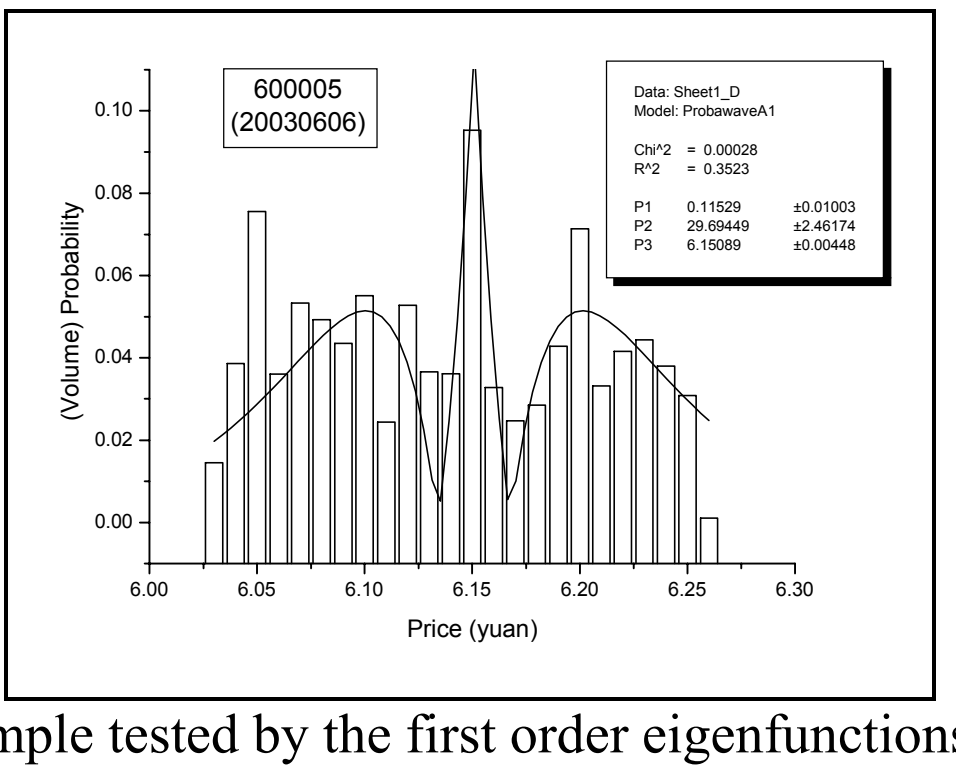

Fig. 6. The sample tested by the first order eigenfunctions at 95% level ($R^2 = 0.35 > R^2_{crit} = 0.16$)

Combining test results in this section and those in Section 2, we conclude that whereas trading price waves around an equilibrium price constantly, the equilibrium price itself characterizes a step or jump change from time to time. An eigenvalue, together with an equilibrium price, determines a stationary state. In addition, price random characteristic is an extremely conditional case in our model. It is the case that its transaction energy or the magnitude of restoring force is a constant over a trading price range—a conserved energy system in a physics perspective. It is rare ($\frac{1}{618} = 0.16\%$ in this paper).

*6.2. Discussions*

In this subsection, we mainly discuss transaction momentum, two sets of analytical solutions, and transaction energy hypothesis (together with methodology).

*6.2.1. Transaction momentum (the volume liquidity) and transaction force*

According to equation (19), transaction momentum does be transaction volume liquidity at a price. In stock market, there is no price change at all if there is no actual trading (volume). The volume plays an exclusive role in determining price change and its liquidity controls the rise or drop of an equilibrium price by its weight. This can explain why large price fluctuations occur when demand is very small [14]. When demand is very small, a small trading volume can produce big impact on its equilibrium price, and even cause its step or jump change. The market is much more unstable.

In addition, transaction momentum rate or transaction volume acceleration is transaction force. The force is defined by equation (20). It describes how we make trading volume, namely, trading behavior in stock market.

A skeptical reader may argue why transaction volume acceleration is not acceleration. Of course, one may define the volume acceleration to be acceleration rather than the force. Let us see what happen if we do so.

1) The transaction volume acceleration is acceleration if and only if we choose the volume as an independent generalized coordinate. In this way, equation (8) is not a holonomic constrain in the study of volume and price because it includes generalized velocity. Not all generalized force can be derivable by differentiating (to the volume) function (s) in this non-holonomic constraint system [32].
2) If the volume is chosen as an independent generalized coordinate, we would study price as a function of the volume. It would be another research topic. Osborne [36], a pioneer in econophysics, observed that price as a function of volume does not exist empirically, and then explained why volume is the function of price, and this is not invertible. McCauley [37] showed that price as a function of volume does not exist mathematically.
3) If the volume is chosen as an independent generalized coordinate, we probably have to study the correlation between the volume and price. In this approach, no analytical solution can be obtained.
4) If the volume is chosen as an independent generalized coordinate, it would be inconvenient for us to observe the volume-price behavior because we choose price as an independent coordinate in financial market.

Thus, it is reasonable for us to choose price rather than volume as an independent variable. The transaction volume acceleration is transaction force rather than acceleration in equation (4).

*6.2.2. Equation solutions*

In Section 5, we get two sets of analytical solutions from transaction volume-price probability wave equation (24). One is a set of zero-order Bessel eigenfunctions (25) if equation (26) is satisfied. In this scenario, transaction energy is not a constant over a trading price range. It is a dissipative energy system in a physics perspective. However, there is coherence between transaction force and restoring force. It is the coherence that the sum of both is equal to an eigenvalue constant over a trading price range. From here, we find that our observed volume distribution pattern is the consequence of a kind of coherence. In Section 2, our empiric test has already demonstrated that equation (26) is satisfied in great majority situations. The volume-price behaves like a probability wave.

The other is a set of eigenfunctions (28) if transaction energy or the magnitude of restoring force is a constant over a trading price range. Under such a condition, restoring force is independent on transaction force. There is no coherence. The function can describe volume uniform distribution or price random walk characteristic.

In physics, the square of a wave function modulus has the meaning of a probability (density). However, we show that it is not true in this model. The volume distribution kurtosis does not behave as steep as the square of a wave function modulus does. A tentative explanation is that function (27) is the consequence of coherence already.

*6.2.3. Methodology and transaction energy hypothesis*

Unlike traditional approaches, we take a transaction system as a “black box” in this paper. What we need to know is input (total amount of transaction) and output (the volume distribution over a trading price range) regardless of what really happens inside the box. We study accumulative or collective trading behavior rather than we plunge ourselves into dark to study individual trading behavior. Actually, we are unable to know exact price path by a one-dimensional problem in a trading system because of separate individual buying and selling decisions in a fully competitive stock market.

Equation (7) is a pure mathematical definition on which a universal theory can be based. When we recast it to transaction energy hypothesis, equation (8), it has some economic and financial meanings. First, when transaction energy is given, its dynamic energy is known, and then, potential energy can be determined by the hypothesis. On the other hand, dynamic energy (distribution energy) can be determined if transaction energy and potential energy are given. We actually propose what dynamic energy is supposed to be over a trading price range in a given potential field when transaction energy is known. By this hypothesis, we study the volume liquidity nonlinear distribution after we choose a linear potential to represent for a restoring force in stock market.

Second, the hypothesis is a holonomic constraint at price coordinate. When we choose price as an independent variable, we simplify an apparently two-dimensional problem (volume and price) to one-dimensional. Otherwise, if we had considered price as a function of time, we would have had to study the volume and price separately rather than whole. For example, Ausloos and Ivanova [15], and Wand and Pandey [16] actually searched for the correlation between the volume and price when a generalized velocity is used. In addition, all the generalized forces are derivable by differentiating function(s) in a holonomic constraint system [32].

Third, one may confuse that price is static in equations (4), (7), and (8) because it is not a function of time. It is not true. Price fluctuation or dynamic behavior does not depend on time but depend on actual trading, i.e. effective supply or demand. There is no price change at all if there is no actual trading in stock market no matter how long it takes. Equations (4), (7), and (8) express transaction energy states at a price. Specifically, equation (8) is a differential expression at a price. Price in the equation is dynamic and not static although it is not a function of time. Does a dynamic variable have to be a function of time?

Forth, with regard to a question on supposition, equation (13), it is clear that the equation is equivalent to transaction energy hypothesis, equation (8), if we assume that the action in equation (11) describes the volume-price behavior or equation (16) is satisfied. Can we suppose that the volume function be expressed by a wave function, equation (11)? Actually, we can also derive the same result as that, equation (24), if we assume that the volume function is an exponent function instead of a wave function. Schŏrdinger [34] got particle wave equation assuming that particle behavior satisfies an exponent function rather than a wave function. Therefore, our supposition is equation (8), together with equation (11) in which the action describes volume-price behavior in a given time interval or transaction amount liquidity behavior. There is no need for any an extra supposition at all.

Finally, we examine our hypothesis by its analytical solutions, equations (1 or 27) and (30). By empirical test, we demonstrate that it hold true at this early stage. In my opinion, most interesting and incredible is probably the hypothesis in this paper. For example, can we extend to derive

Galaxy density distribution in astrophysics [38] or electron density distribution in double-slit electron interference experiment in quantum physics from a similar hypothesis?

### *6.3. Possible applications*

There are many possible applications using this effective supply-demand volume-price wave model. For example, we understand that transaction momentum or volume liquidity plays an exclusive role in determining the rise or drop of an equilibrium price through its weight in a transaction system. We can estimate both risk and opportunity on a market by the momentum. In addition, it is convenient for us to model an entire price volatility distribution very well in which fat tail and cluster exist, and forecast the volatility probability. Trading institutions, moreover, can use the ratio of cash (demand quantity) to the market value of stocks (supply quantity) they hold to measure the risk or opportunity they have in financial market. Whereas it is difficult to distinguish between current behavioral theory and traditional rational theory because of mathematical and predictive similarities [39], furthermore, this model provides a new methodology to study behavioral finance quantitatively because of measurable coherence between transaction force and restoring force. With this model, we are able to simulate actual trading in stock market. For example, how much amount of money are we going to use to trade a stock price from current $5 dollar per share to expected $10 dollar per share in a speculative strategy, or to boost Shanghai Stock Index from present 1000 points to prospect 1500 points in the market strategy that accelerates transaction liquidity, avoids crisis, and makes the market profitable?

## 7. Summaries and conclusions

We test our observation that there is a stationary transaction volume distribution over a trading price range on individual stocks in an intraday trading time interval. It results that the volume has a main kurtosis near the price mean value in great majority situations, and concludes that our observation holds true. In order to explain the volume-price behavior, we propose a transaction energy hypothesis and find the linear potential that can represent for effective supply-demand restoring force. Based on both, we derive a time-independent security transaction volume-price probability wave equation and get two sets of analytical transaction volume distribution eigenfunctions over a trading price range. One is a set of zero-order Bessel eigenfunctions (25) if there is coherence. From this solution, we find that our observed volume distribution pattern is the consequence of a kind of coherence. The other is a set of eigenfunctions (28) if transaction energy or the magnitude of restoring force is a constant over a trading price range. It can describe the volume uniform distribution over a trading price range or price random walk characteristic. By empiric test, we demonstrate the model validation at this early stage. The volume-price behaves like a probability wave.

## Acknowledgement

I especially appreciate Suhua Liao, my mother, for understanding and supporting my career constantly.

**Appendix A: Derivation of a time-dependent volume-price probability wave equation**

Suppose that a transaction volume-price probability wave function $\psi(p)$ is

$$\psi(p) = R \cdot e^{iS/B}, \tag{A-1}$$

where $R$ is the wave amplitude, $S$ its action or Hamiltonian principal function, and $B$ a constant to make its phase dimensionless [34,35].

From equation (A-1), we have

$$\frac{\partial S}{\partial p} = -\frac{iB}{\psi}\frac{\partial \psi}{\partial p} \tag{A-2}$$

and

$$\frac{\partial S}{\partial t} = -\frac{iB}{\psi}\frac{\partial \psi}{\partial t}. \tag{A-3}$$

Assume that the volume-price behavior satisfies a Hamilton-Jacobi equation in this interacting energy system as

$$\frac{\partial S}{\partial t} + H(p, \frac{\partial S}{\partial p}) = 0, \tag{A-4}$$

where we construct a Hamiltonian that is equal to the sum of transaction dynamic energy and potential energy. It is written as

$$H(p, \frac{\partial S}{\partial p}) = T\left( p, \frac{\partial S}{\partial p} \right) + W(p). \tag{A-5}$$

In equation (A-5), dynamic energy and potential energy are constrained by transaction energy hypothesis, equation (8).

Let us consider two scenarios in a transaction system. If we assume

$$\frac{\partial S}{\partial t} = -E \tag{A-6}$$

or

$$iB\frac{\partial \psi}{\partial t} = E\psi, \tag{A-7}$$

then, we get a time-independent equation (24). In equation (A-6), we suppose that the action in function (A-1) describe volume-price behavior or transaction amount in a given time interval. Otherwise, we use equation (A-3) and (19), and write equation (A-4) as

$$-i\frac{B}{\psi}\frac{\partial \psi}{\partial t} + \frac{p}{V}\left(\frac{\partial S}{\partial p}\right)^2 + W(p) = 0. \tag{A-8}$$

Making due allowance for the complex nature of $\psi$, we recast equation (A-8) as

$$-i\frac{B}{\psi}\frac{\partial \psi}{\partial t} + \frac{p}{V}\left(\frac{\partial S}{\partial p}\right)^* \left(\frac{\partial S}{\partial p}\right) + W(p) = 0. \tag{A-9}$$

We construct a transaction energy functional $G(p,\psi)$ using equation (A-2) as follow:

$$G(p,\psi) = \left[ -i\frac{B}{\psi}\frac{\partial \psi}{\partial t} + W(p) \right]\psi\psi^* + \frac{B^2}{V}p\left(\frac{\partial \psi^*}{\partial p}\right)\left(\frac{\partial \psi}{\partial p}\right). \tag{A-10}$$

Using equation (23), we thus write a time-dependent volume-price wave equation as

$$-i\frac{V}{B}\frac{\partial \psi}{\partial t} = p\frac{\partial^2 \psi}{\partial p^2} + \frac{\partial \psi}{\partial p} - \frac{V}{B^2}W(p)\psi. \tag{A-11}$$

Obviously, if equation (A-7) is satisfied, then, equation (A-11) is a time-independent equation (24).

**Appendix B**

Given a differential equation

$$\frac{d^2\psi}{dp'^2}+\frac{1}{p'}\frac{d\psi}{dp'}+\left(\frac{E}{p'}-A\right)\psi=0 \qquad \text{(B-1)}$$

and natural boundary conditions

$$\psi(0)<\infty \text{ and } \psi(\pm\infty)\to 0, \qquad \text{(B-2)}$$

where $E$ is a constant and $p'=p-p_0$, find its solution.

In equation (B-1), there is a regular singular point at $p'=0$ and non-regular singular points at $p'=\pm\infty$.

For $p'\geq 0$:

If $p'\to 0$, then, let

$$\psi_0(p')\propto p'^{\rho}. \qquad \text{(B-3)}$$

Substituting (B-3) into (B-1), we have identical equation as

$$\rho(\rho-1)+\rho=0$$

or $\rho=0$. (B-4)

So, we write

$$\psi_0(p')\propto const. \qquad \text{(B-5)}$$

If $p'\to+\infty$, then, the equation (B-1) is approximately as

$$\frac{d^2\psi_{+\infty}(p')}{dp'^2}-A\,\psi_{+\infty}(p')=0. \qquad \text{(B-6)}$$

Because of eigenvalue $A>0$, the asymptotic equation (B-6) has a solution as

$$\psi_{+\infty}(p')\sim e^{\pm\sqrt{A}p'}. \qquad \text{(B-7)}$$

According to natural boundary condition $\psi(+\infty)\to 0$, only $\psi_{+\infty}(p')\sim e^{-\sqrt{A}p'}$ is acceptable.

Assume that a general solution for equation (B-1) is

$$\psi(p')=\psi_0(p')\cdot\psi_{+\infty}(p')\cdot u(p')\propto e^{-\sqrt{A}p'}\cdot u(p'). \qquad \text{(B-8)}$$

Substituting the trial solution (B-8) back into equation (B-1), we yield

$$\frac{d^2u(p')}{dp'^2}+\left(\frac{1}{p'}-2\sqrt{A}\right)\frac{du(p')}{dp'}+\left(\frac{E-\sqrt{A}}{p'}\right)u(p')=0. \qquad \text{(B-9)}$$

Introducing a new variable $\xi=2\sqrt{A}p'$ into the equation above, we produce

$$\xi\frac{d^2u(\xi)}{d\xi^2}+(1-\xi)\frac{du(\xi)}{d\xi}-\left(\frac{1}{2}-\frac{E}{2\sqrt{A}}\right)u(\xi)=0, \qquad \text{(B-10)}$$

Comparing equation (B-10) with a confluent hypergeometric equation

$$\xi\,\frac{d^2u(\xi)}{d\xi^2}+(\gamma-\xi)\frac{du(\xi)}{d\xi}-\alpha u(\xi)=0, \qquad \text{(B-11)}$$

we have

$$\gamma=1 \text{ and } \alpha=\frac{1}{2}\left(1-\frac{E}{\sqrt{A}}\right). \qquad \text{(B-12)}$$

The analytical solution near the origin $\xi\approx 0$ for equation (B-11) is the confluent hypergeometric function or the first Kummer's function $F(\alpha,\gamma,\xi)$, where

$$F(\alpha,\gamma,\xi)=\sum_{k=0}^{\infty}\frac{(\alpha)_k}{k!(\gamma)_k}\xi^k,\ (\gamma\neq 0 \text{ and not a negative integer}) \qquad \text{(B-13)}$$

$$(\alpha)_k = \frac{\Gamma(\alpha + k)}{\Gamma(\alpha)}, \tag{B-14}$$

$$\Gamma(x) = \frac{\Gamma(x+1)}{x}. \tag{B-15}$$

The function $F(\alpha, \gamma, \xi)$ has the important property that, if $\alpha = -m(m = 0,1,2,3\cdots)$, it reduces to a polynomial of order $m$ so as to satisfy natural boundary condition. Hence, the eigenvalues $A_m$ from (B-12) are

$$\sqrt{A_m} = \frac{E}{1+2m}, \ (E > 0) \tag{B-16}$$

or

$$A_m = \frac{E^2}{(1+2m)^2}. \tag{B-17}$$

Thus, the solution for equation (B-9) is

$$u_m(p') = F[-m,1,\xi_m] = F[-m,1,2\sqrt{A_m}p']. \tag{B-18}$$

Substituting the solutions (B-18) into (B-8), we have the eigenfunctions $\psi_m(p')$ for equation (B-1) as,

$$\psi_m(p') = C_m e^{-\sqrt{A_m}p'} \cdot F\left(-m,1,2\sqrt{A_m}p'\right), \tag{B-19}$$

where $C_m$ is a normalized constant, $m = 0,1,2,3,\cdots$

If $p' \le 0$, Let $p' = -|p'|$, then, the equation (B-1) is

$$\frac{d^2\psi}{d|p'|^2} + \frac{1}{|p'|}\frac{d\psi}{d|p'|} + \left(-\frac{E}{|p'|} - A\right)\psi = 0. \tag{B-20}$$

Using the same method, we can derive the solution as

$$\sqrt{A_m} = -\frac{E}{1+2m} = \frac{|E|}{1+2m}{}^{*}, \tag{B-21}$$

or

$$A_n = \frac{E^2}{(1+2m)^2}, (E = -|E| < 0). \tag{B-22}$$

Their eigenfunctions for equation (B-20) are

$$\psi_m(p') = e^{-\sqrt{A_m}|p'|} \cdot F\left(-m,1,\sqrt{A_m}|p'|\right) \tag{B-23}$$

Combining (B-19) and (B-23), we have a general solution $\psi(p')$ for equation (B-1) over the range $-\infty < p' < +\infty$ as

$$\psi_m(p') = e^{-\sqrt{A_m}|p'|} \cdot F\left(-m,1,2\sqrt{A_m}|p'|\right), \tag{B-24}$$

with $$A_m = \frac{E^2}{(1+2m)^2}. \tag{B-25}$$

where $m = 0,1,2,3,\cdots$; $E > 0$ if $p' \ge 0$, and $E = -|E| < 0$ if $p' \le 0$.

---

[*] Since $E = p' v_{tt}$, $E = -|E| < 0$ holds if $p' \le 0$